\documentstyle[aps,epsf]{revtex}

\begin{document}  

\baselineskip 14pt
\title{Extruded Plastic Scintillation Detectors}
\author{Anna Pla-Dalmau, Alan D. Bross and Kerry L. Mellott}
\address{Fermi National Accelerator Laboratory\\
 P. O. Box 500, Batavia,         IL 60510, USA}

\maketitle

\begin{abstract}
As a way to lower the cost of plastic scintillation detectors,
commercially available polystyrene pellets have been used in
the production of scintillating materials that can be extruded
into different profiles.  The selection of the raw materials
is discussed.  Two techniques
to add wavelength shifting dopants to polystyrene pellets and to
extrude plastic scintillating strips are described.
Data on light yield and transmittance measurements are presented.

\end{abstract}

\section{Introduction}              

Plastic scintillation detectors have been used in nuclear and
high energy physics for many decades
\cite{Birks}.
Their advantages and disadvantages are recognized.  Among their
benefits are fast response, ease of manufacture and versatility.
Their main drawbacks are radiation resistance and cost.
Many research projects have concentrated on improving the
fundamental properties of plastic scintillators
\cite{SCIFI93,SCIFI97}, but little
attention has focussed on their cost.  Currently
available plastic scintillating materials are high quality
products whose cost is relatively expensive, and because of
that, their use in very large detectors has not been a
feasible option.
For instance, 
MINOS (Main Injector Neutrino Oscillation Search)
will require 400,000 Kg of plastic scintillator for its
detector
\cite{MINOS}.  With the price of cast
scintillator at approximately \$40 per Kg, such a detector would not
be affordable.  However, 
recent studies using
commercial polystyrene pellets as the base material for extrudable
plastic scintillators have allowed the MINOS collaboration
to consider a less expensive alternative in building
a plastic scintillation detector.  Furthermore, the D0 experiment at
Fermilab has been able to use extruded plastic scintillator to
build and upgrade their Forward and Central Preshower Detectors.  
In this case,
the driving force was not the cost of the material,
since only 2,000 Kg of plastic
scintillator were needed, but the opportunity to use a particular
shape (triangular bar)
that would have been expensive to machine out of cast plastic
scintillator sheets
\cite{D01,D02}.

\section{Extruded Plastic Scintillators}

Several factors contribute to the high cost of plastic
scintillating sheets and wavelength shifting fibers.  
The main reason is 
the labor-intensive nature of the manufacturing process.  
The raw materials, namely styrene, vinyltoluene, and the
dopants, need to be highly pure.  
These purification steps often take place just prior to the
material utilization.
Cleaning and assembly of the molds for the polymerization process
is a detail-oriented operation that adds to the overall timeline.
The polymerization cycle lasts several days.  It consists of a high
temperature treatment to induce full conversion from monomer
to polymer, followed by a controlled ramp-down to room temperature to
achieve
a stress-free material.
Finally, there are machining charges for sheets and tiles,
and drawing charges for fibers that cannot be overlooked.

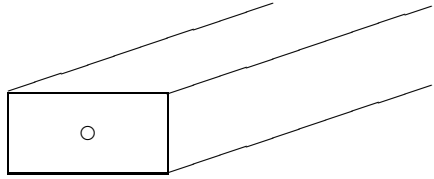
\begin{figure}[]
\begin{picture}(432,108)
\put(144,36){\framebox(60,30)}
\put(174,51){\circle{5}}
\put(144,67){\line(3,1){100}}
\put(204,36){\line(3,1){100}}
\put(204,66){\line(3,1){100}}
\end{picture}
\vskip -.5 cm
\caption[]{
\label{rect}
\small Rectangular scintillator profile with a hole in the middle
for a WLS fiber.}
\end{figure}

In order to significantly lower the cost of plastic scintillators,
extruded plastic scintillation materials need to be considered.
In an extrusion process, polymer pellets or powder must
be used.  Commercial polystyrene pellets are readily available,
thus eliminating monomer purification and polymerization charges.
In addition, extrusion can manufacture nearly any shape,
increasing detector geometry options.  There are, however, some
important disadvantages.  The extruded plastic scintillator is
known to have poorer optical quality than the cast material.
The main cause is high particulate matter content
within the polystyrene pellets.  General purpose
polystyrene pellets are utilized in numerous products but
few of them have strict optical requirements.
A way to bypass the short attenuation length problem is to extrude
a scintillator shape and use a wavelength shifting (WLS) fiber
as readout.  Our first approach was a two-step process
that involved adding dopants to commercial polystyrene pellets to
produce scintillating polystyrene pellets, which were then used
to extrude a scintillator profile with a hole in the middle
for a WLS fiber (Figure~\ref{rect}).
The goal in the first step was to prepare scintillating pellets
of acceptable optical quality in a factory environment.
In addition to careful selection of the raw materials, the
manufacturing concerns dealt with possible
discoloration of the scintillating pellets because of
either residues present in the equipment or degradation
of the polymer pellets and the dopants in the processing device.
The latter could be induced by the presence of oxygen
at the high temperatures and pressures which constitute
the typical operating conditions.

After producing the first batch of scintillating pellets, samples
were cast to perform light yield and radiation
degradation studies.
Samples of standard cast scintillators such
as BC404 and BC408, and samples prepared through bulk polymerization
at Fermilab were also included in the studies (Table I).  
All the samples had similar dopant composition and were cut as 2-cm
cubes.  The light yield measurements were performed using a
$^{207}$Bi source (1 MeV electrons).
The light yield results (Table I) showed no significant
difference among them.  The Bicron samples are made of 
poly(vinyltoluene) instead of polystyrene which accounts
for the 20\% increase in light output \cite{Birks}.
The samples for radiation damage studies were placed in stainless
steel cans and connected to a vacuum pump for two weeks to remove
dissolved air and moisture.  The cans were then
back-filled with nitrogen and irradiated with a $^{60}$Co source
at the Phoenix Memorial Laboratory of the University of Michigan.
The irradiations took place
at a rate of approximately 15 KGy/h to a total dose of 1 KGy.
After irradiation and annealing, 
the extruded scintillator cubes showed a 5\%
decrease in light yield which is similar to the losses observed
in regular scintillator of this composition.
Based on these tests, there was no sign of degradation in the
scintillating pellets.  The material was then used to produce extruded
scintillator of different profiles with a hole in the middle
for a WLS fiber.

\vspace{12pt}
\begin{center}
\small{
TABLE I.  Relative light yield of samples with similar compositions
but from different manufacturing processes.}\\
\begin{tabular}{cccc}                \hline\hline
\mbox{{\hspace{0.5in}}Scintillator{\hspace{0.5in}}} &
\mbox{{\hspace{0.5in}}Bicron$^{a}${\hspace{0.5in}}} &
\mbox{{\hspace{0.5in}}extruded{\hspace{0.5in}}} &
\mbox{{\hspace{0.5in}}bulk polymerized{\hspace{0.5in}}} \\
\hline
404&1.0&0.80&0.78\\
408&1.0&0.85&0.77\\
\hline\hline
\end{tabular}
\noindent
$^{a}$Bicron scintillator has a poly(vinyltoluene) matrix which
yields 20\% more light than a polystyrene one \cite{Birks}.
\end{center}

\subsection{Selection of Raw Materials}

\begin{figure}[]      % in second brace, h=here, t=top, b=bottom
\centerline{\epsfxsize 3.5 truein \epsfbox{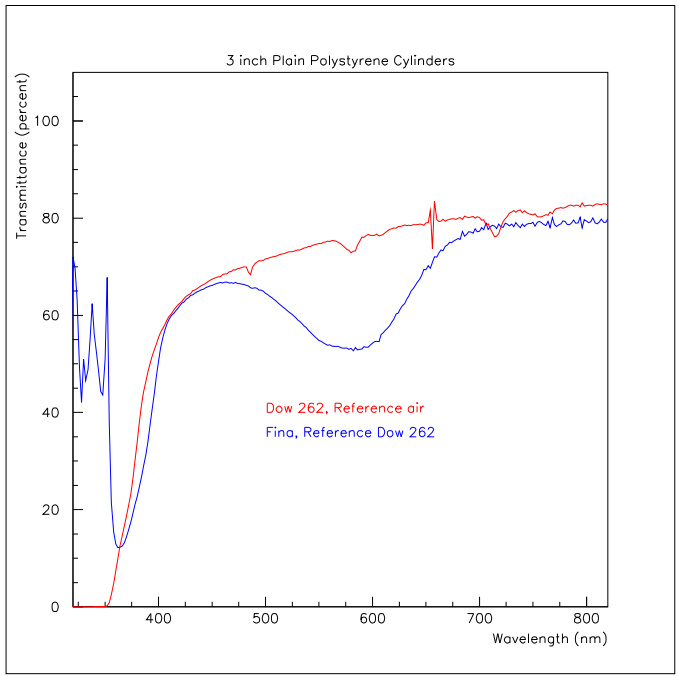}}
\vskip 0.2 cm
\caption[]{
\label{trans}
\small Transmittance data of commercial polystyrene pellets.}
\end{figure}

There are many manufacturers and grades
of polystyrene pellets.  Most of them fall under the category of general
purpose polystyrene.  Only a few offer optical
quality polystyrene pellets.  Needless to say,
there is a substantial difference in price.  Nonetheless,
the first plastic scintillating pellets were prepared using
an optical grade polystyrene from Dow, labeled XU70251, which
was later superseded by XU70262 (Dow 262).
The price for Dow 262 is about \$4.5 per Kg.
After confirmation by the initial tests that high quality extruded plastic
scintillators were feasible, the quest began to replace the
costly optical grade pellets with general purpose material.
Various samples of different polystyrene grades were received
from Dow, Fina, Nova, BASF, Huntsman, etc.  These samples had been
selected based on price, availability and melt flow rate for
ease of extrusion.  These materials were cast into cylinders up to
3 inches long.  Transmittance measurements were performed using
a Hewlett-Packard 8452 spectrophotometer.  The materials tested
were compared to cast samples of Dow 262 pellets.  
Often polystyrene contained additives that absorbed at long wavelengths
such as the Fina pellets illustrated in
Figure~\ref{trans}.  This absorption would diminish the amount
of light produced by the dopants that need to be added
to make a particular scintillator.
Other features observed were long absorption tails and haziness caused by
additives and debris in the pellets.
There were a couple of materials that were repeatedly tested
and showed high clarity and lack of absorptions at long wavelengths.  
Dow Styron 663 (Dow 663) was chosen
as the general purpose polystyrene grade to conduct our extrusion
studies.  Its price ranges from \$1.3/Kg to \$1.7/Kg depending, among
other things, on the quantity ordered.

\begin{figure}[t]      % in second brace, h=here, t=top, b=bottom
\centerline{\epsfxsize 3.5 truein \epsfbox{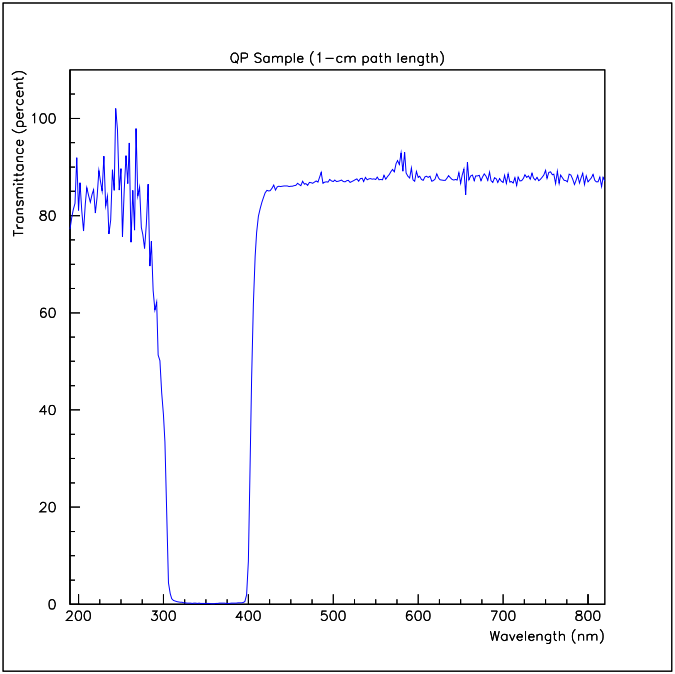}}
\vskip 0.2 cm
\caption[]{
\label{ppo}
\small Transmittance data of extruded plastic scintillator.}
\end{figure}

A variety of organic fluorescent compounds can be used
as primary and secondary dopants in plastic scintillator
applications.
The primary dopant is commonly used at a 1--1.5\% (by weight)
concentration.  The secondary dopant or wavelength shifter 
in scintillator is
utilized at a concentration of 0.01--0.03\% (by weight).
The goal was to prepare a blue-emitting scintillator that could
be readout with a green WLS fiber.  Most green fibers are doped
with K27, and thus the emission of the scintillator would have to match
as best as possible the absorption of K27 in the fiber.
The selection of dopants was based on these spectroscopic 
requirements as well as price and ease of manufacture.
{\em para}-Terphenyl (\$200--225/Kg) and PPO (\$100--160/Kg) were
considered as primary dopants.  
POPOP and bis-MSB (both at \$0.5--1/g) were
tested as secondary dopants.  The final choice for the extruded plastic
scintillator was PPO and POPOP in Dow 663.
Figure~\ref{ppo} plots the transmittance spectrum of an extruded
scintillator sample of this type.

\subsection{Manufacturing Techniques}

\begin{figure}
\begin{picture}(432,288)
\put(30,255){\framebox(120,30){polystyrene pellets}}
\put(30,215){\framebox(120,30){dopants and additives}}
\put(200,235){\framebox(120,30){mixture batch}}
\put(270,195){\framebox(120,30){inert gas (argon)}}
\put(200,145){\framebox(120,30){processing device}}
\put(364,145){\framebox(90,30){extrusion}}
\put(320,160){\vector(1,0){44}}
\put(230,235){\vector(0,-1){60}}
\put(290,195){\vector(0,-1){20}}
\put(170,270){\line(0,-1){40}}
\put(170,250){\vector(1,0){30}}
\put(150,270){\line(1,0){20}}
\put(150,230){\line(1,0){20}}
\thicklines
\put(200,85){\framebox(120,40){\bf scintillating pellets}}
\put(160,15){\framebox(200,50){\bf extrude to form scintillating profile}}
\put(260,85){\vector(0,-1){20}}
\put(260,145){\vector(0,-1){20}}
\thinlines
\put(360,92){\framebox(80,25){cast}}
\put(80,92){\framebox(80,25){injection mold}}
\put(320,105){\vector(1,0){40}}
\put(200,105){\vector(-1,0){40}}
\end{picture}
\caption[]{
\label{meth1}
\small Two-step process: batch coloring and extrusion (Method 1).}
\end{figure}
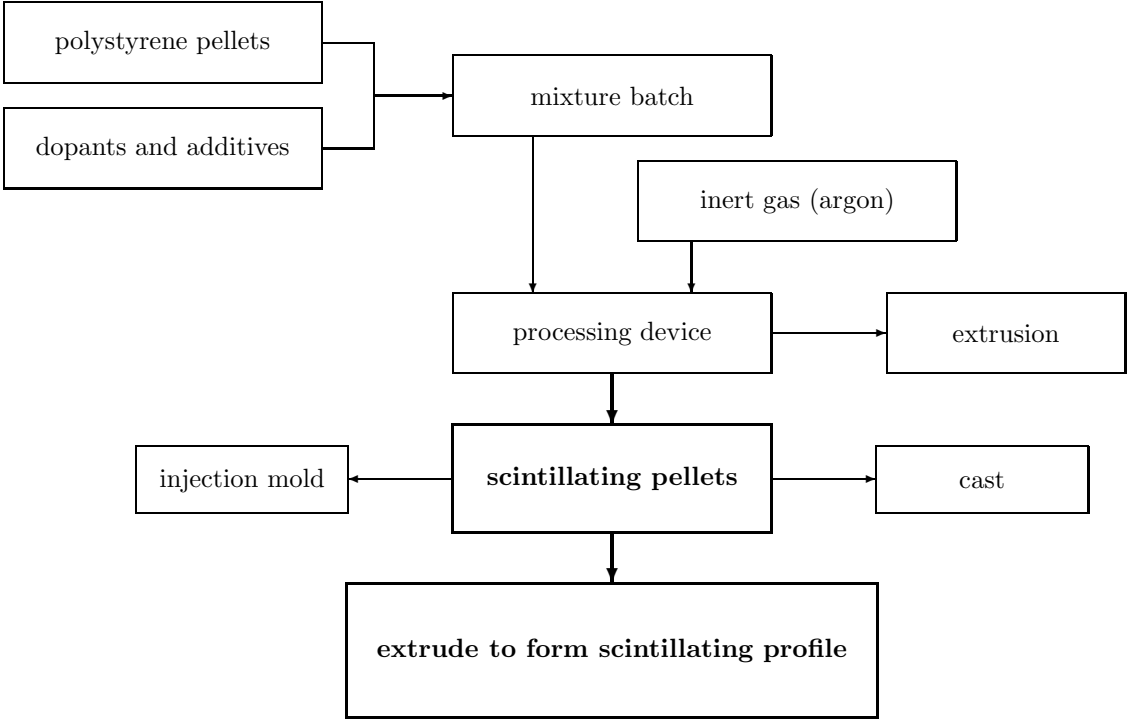

The majority of the extruded scintillator prepared has used 
Method 1, a
two-step process conducted at two separate facilities.
Figure~\ref{meth1} depicts the flow chart for this method.
The first step was carried out at a company whose
function was to add the dopants to the polystyrene pellets.
(In the plastics industry, this trade is typically referred
to as a color or compounding business.)
Prior to the coloring run, polystyrene pellets were purged for several
days with an inert gas, generally argon, to remove dissolved oxygen
and moisture.  The coloring step was a batch process where polystyrene
pellets and dopants were tumble-mixed for 15 min.\ and then added to the
hopper of an extruder.  Each batch prepared 45 Kg of mixture.
A silicone oil was used as a coating aid to achieve better distribution
of the dopants on the pellet surface.
An argon flow was also added to the hopper to minimize the presence of
oxygen in the extruder.  The die at the extruder head generated several
strings of material which were cut yielding the scintillating
pellets.  At the end, the scintillating pellets 
collected in many containers during
the run were blended to homogenize the material.
These pellets could now be used to produce plastic scintillators through
several procedures --- namely extrusion, casting and injection molding.
In this case, the scintillating pellets were
taken to an extrusion company to extrude the desired scintillator profile.

Using this batch process, there is also the possibility of directly
extruding the scintillator profile and thus \mbox{by-passing}
the pelletizing
step.  This is the route that the MINOS collaboration has chosen to
investigate.  This variation of Method 1 can be less expensive since
all the work is done in one facility.  It also reduces the heat history
of the product by removing its exposure to another high temperature
cycle and minimizes the chance of optical degradation.
The drawback is in the batch work since the polymer and the dopants
still need to be weighed for each mixture, and in the tumble-mixing step
which is susceptible to contamination and prone to errors.

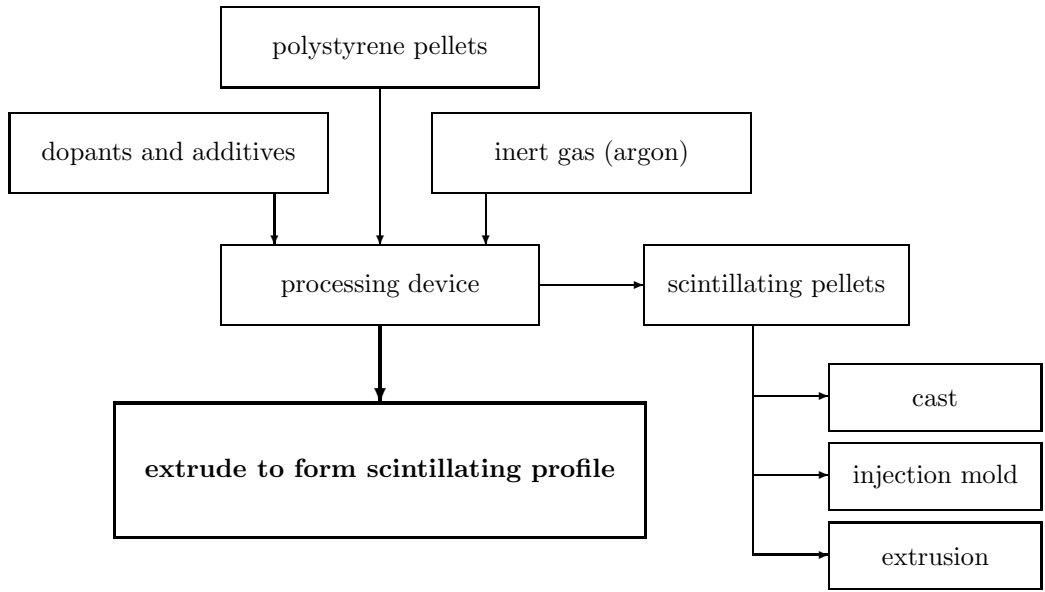
\begin{figure}
\begin{picture}(432,288)
\put(144,250){\framebox(120,30){polystyrene pellets}}
\put(64,210){\framebox(120,30){dopants and additives}}
\put(224,210){\framebox(120,30){inert gas (argon)}}
\put(144,160){\framebox(120,30){processing device}}
\put(304,160){\framebox(100,30){scintillating pellets}}
\put(374,120){\framebox(80,25){cast}}
\put(374,90){\framebox(80,25){injection mold}}
\put(374,60){\framebox(80,25){extrusion}}
\thicklines
\put(104,80){\framebox(200,50){\bf extrude to form scintillating profile}}
\put(204,160){\vector(0,-1){30}}
\thinlines
\put(204,250){\vector(0,-1){60}}
\put(164,210){\vector(0,-1){20}}
\put(244,210){\vector(0,-1){20}}
\put(264,175){\vector(1,0){40}}
\put(345,160){\line(0,-1){87}}
\put(345,133){\vector(1,0){29}}
\put(345,103){\vector(1,0){29}}
\put(345,73){\vector(1,0){29}}
\end{picture}
\vskip -1.7 cm
\caption[]{
\label{meth2}
\small Continuous in-line coloring and extrusion process (Method 2).}
\end{figure}

An alternative to these operations is given by Method 2
which is summarized in
Figure~\ref{meth2}.  Method 2 is a continuous in-line coloring and
extrusion process.  It emphasizes the most direct pathway from
polystyrene pellets to the scintillator profile with the least
handling of raw materials.
In this situation, the purged polystyrene pellets and dopants are
metered into the extruder at the correct rate for the required
composition of the scintillator.  An argon flow is still used
at the hopper.  Coating agents are no longer needed.
The appropriate die profile gives rise to the
extruded scintillator form of choice.  If the die can produce
strands, these can also be pelletized and the scintillating
pellets used in other processes.
Method 2 has been tested and produces plastic scintillator of
high quality and homogeneity.
Although it is a simple concept, the equipment needed to accurately
meter small quantities of powders such as the dopants and
to achieve a good distribution of the powders in the molten
polymer is not widely available.  The difficulty in testing this
process was finding a facility with the adequate instrumentation.

\section{Light Yield of Extruded Plastic Scintillators}

Light yield studies have been performed on many samples of extruded
plastic scintillators.  Although a variety of shapes and sizes
is available, the measurements have mostly been carried out on
11.5-cm long rectangular extrusions (1cm x 2cm) with a hole in the
middle for a green WLS fiber.  
Each extrusion is tightly wrapped in Tyvek for this test.
The WLS fiber utilized is BC91A
(0.835 mm diameter, 1.5 m long) with one mirrored end.
The light yield test setup uses an electron spectrometer with
$^{106}$Ru source whose 3-MeV beam is momentum selected.
There is a small trigger counter in front of the extruded sample.
The photomultiplier tube used is a Hamamatsu R2165
which has excellent single photo-electron resolution.
The fiber is held at a fixed position from the PMT surface
to minimize fluctuations among measurements.
The light yield is determined from the following calculation:

\[Light Yield = \frac{Mean-Pedestal}{Gain}\]

\begin{figure}[]      % in second brace, h=here, t=top, b=bottom
\centerline{\epsfxsize 3.5 truein \epsfbox{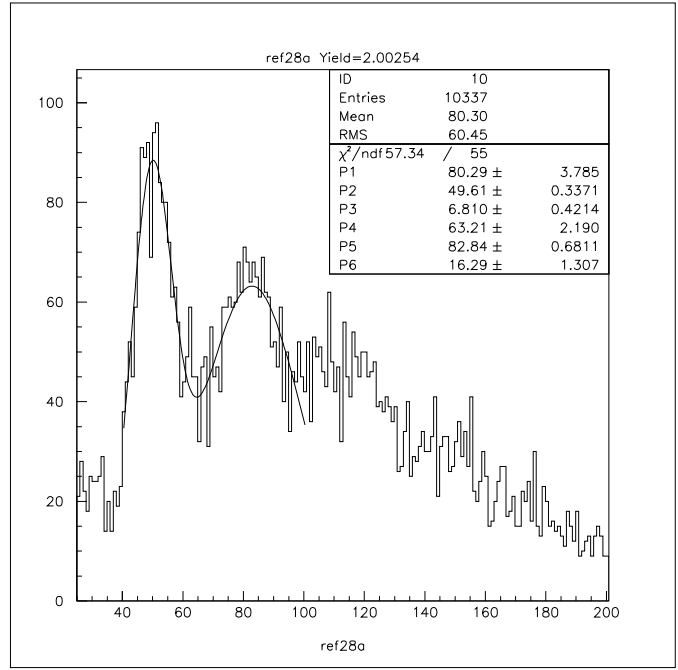}}
\vskip 0.2 cm
\caption[]{
\label{light}
\small Light yield distribution of extruded plastic
scintillator (RDN 262) with WLS fiber.  Solid curve is 2-gaussian
fit to first and second peaks.}
\end{figure}

\vspace{12pt}
\noindent
where the mean and the gain are defined as:

\[Mean = \frac{\sum_{i}^{n} v_{i} x_{i}}{\sum_{i}^{n} v_{i}}\]

\[Gain = First Peak - Pedestal\]

\noindent 
where v$_{i}$ is the number of entries for each ADC value, x$_{i}$.
The data are fitted to locate the position of the first and second
peaks, and the pedestal.
Figure~\ref{light} presents the light yield distribution of an
extruded sample and the fit for the first and second electron peaks.

The results from a series of light yield measurements are
listed in Table II.
RDN 262 extrusions were prepared by the two-step batch process (Method 1)
using Dow 262 optical grade pellets.
Leistritz 262 and 262P samples were produced by the continuous
procedure (Method 2) using Dow 262 polymer.
Leistritz 663 samples were also prepared by Method 2 but used
general purpose polystyrene pellets (Dow 663).
Although the samples are from different runs, their light output
is similar.  The Leistritz 262 samples show a slightly lower
light yield but their profile is smaller than that of the
remaining samples.  
These samples were collected early in the extrusion run
when the profile was not completely to specification.
These results indicate that there is
no major difference in light yield between Method 1 and Method 2.  
This test proves that the
continuous in-line coloring and extrusion process (Method 2) yields
a homogeneous part with the right concentration of dopants.
This aspect is less of a concern in Method 1 since the first step 
includes batch tumble-mixing and post-blending of the
scintillating pellets before the scintillating profile is extruded.
In addition, these numbers confirm that Dow 663 (general purpose
polystyrene pellets) can replace the optical grade pellets
initially utilized.
More measurements are underway to compare these extrusions 
to samples of commercial plastic scintillator sheets
which have been cut to the same profile.

\vspace{12pt}
\begin{center}
\small{
TABLE II.  Light yield of extruded plastic
scintillator samples.}\\
\begin{tabular}{lcccc}                \hline\hline
\mbox{{\hspace{0in}}Scintillator{\hspace{0.5in}}} &
\mbox{{\hspace{0.25in}}No. of samples{\hspace{0.25in}}} &
\mbox{{\hspace{0.25in}}Light Yield{\hspace{0.25in}}} &
\mbox{{\hspace{0.25in}}St. Dev.{\hspace{0.25in}}} &
\mbox{{\hspace{0.25in}}Characteristics{\hspace{0.25in}}}\\
\hline
RDN 262&30&2.05&0.09&Dow 262, Method 1\\
Leistritz 262&10&1.81&0.14&Dow 262, Method 2\\
Leistritz 262P&10&2.02&0.10&Dow 262, Method 2\\
Leistritz 663&15&2.22&0.07&Dow 663, Method 2\\
\hline\hline 
\end{tabular}
\end{center}

\section{Conclusions}

Research on extruded plastic scintillator was driven
by the high cost of cast plastic scintillator.
The goal was to use commercially available polystyrene
pellets, in particular from a general purpose grade,
and standard extrusion equipment to lower the price
of plastic scintillators.
Extruded plastic scintillator strips have
been manufactured and tested.  
The estimated price for extruded scintillator ranges
from \$3.5/Kg to \$6/Kg.  About 50\% of the cost
is due to raw materials with the remaining 50\%
due to processing.
The results indicate
that the extruded scintillator profile with
a WLS fiber as readout is a valid system for
scintillation detectors.
The MINOS experiment will build a very large detector
using this technology.
The D0 experiment is assembling the Central and Forward
Preshower detectors using extruded scintillating
triangular strips.

\end{document}